\documentclass[twocolumn,amsmath,amssymb,floatfix,prb,showpacs,footinbib,superscriptaddress]{revtex4}
\usepackage{amsmath,amssymb,natbib,bm,graphicx,url,psfrag,epsfig,dsfont}
\usepackage[ansinew]{inputenc}
\usepackage[normalem]{ulem} 
\newcommand{\be}{\begin{equation}}
\newcommand{\ee}{\end{equation}}
\newcommand{\bes}{\begin{equation}\begin{split}}
\newcommand{\ees}{\end{split}\end{equation}}

\newcommand{\vc}[1]{\text{\boldmath{$ #1 $}}}

\newcommand{\mean}[1]{\langle  #1  \rangle}

\newcommand\grad{\vc\nabla}

\newcommand\etal{{\it et al.}}

\begin{document}

\title{Relaxation mechanisms of the persistent spin helix}

\author{Matthias C.\ \surname{L\"uffe}}
\affiliation{Dahlem Center for Complex Quantum Systems and Fachbereich Physik, Freie
Universit\"at Berlin, Arnimallee 14, 14195 Berlin, Germany}

\author{Janik \surname{Kailasvuori}}
\affiliation{Max-Planck-Institut f\"ur Physik komplexer Systeme, N\"othnitzer Stra\ss{}e 38, 01189 Dresden, Germany }

\author{Tamara S. \surname{Nunner}}
\affiliation{Dahlem Center for Complex Quantum Systems and Fachbereich Physik, Freie
Universit\"at Berlin, Arnimallee 14, 14195 Berlin, Germany}

\date{\today}

\begin{abstract}
We study the lifetime of the persistent spin helix in semiconductor quantum wells with equal Rashba- and linear Dresselhaus spin-orbit interactions.
In order to address the temperature dependence of the relevant spin relaxation mechanisms we derive and solve semiclassical spin diffusion equations taking into account spin-dependent impurity scattering, cubic Dresselhaus spin-orbit interactions and the effect of electron-electron interactions. For the experimentally relevant regime we find that the lifetime of the persistent spin helix is mainly determined by the interplay of cubic Dresselhaus spin-orbit interaction and electron-electron interactions. We propose that even longer lifetimes can be achieved by generating a spatially damped spin profile instead of the persistent spin helix state.

\end{abstract}


\maketitle

\section{Introduction}

Within the field of spintronics,\cite{book_awschalom_2002,zutic_2004,awschalom_2007} semiconductor devices with spin-orbit coupling have attracted great attention over the past years because they offer a setting where electronic spin polarizations can be generated and manipulated in the absence of ferromagnetism or external magnetic fields.\cite{awschalom_2009} This opens the perspective of adding the spin degree of freedom to the existing semiconductor logic in information technology without encountering the challenge of artificially integrating local magnetic fields in devices. From this applications point of view it is clearly desirable to maximize the spin lifetimes and coherence lengths in semiconductor spintronics devices. 

In this respect an ideal candidate is the persistent spin helix (PSH), a spin density wave state with infinite lifetime, which exists in two-dimensional electron systems with Rashba and linear Dresselhaus spin-orbit interaction of equal magnitude~\cite{schliemann,bernevig1} due to a SU(2) symmetry of the corresponding Hamiltonian.~\cite{bernevig1} On a less abstract level this can be understood as the combined effect of diffusion and spin precession: the momentum-dependent spin-orbit field is perpendicular to the PSH wave vector, and its magnitude grows linearly with the projection of the momentum argument on the direction of this wave vector. If, for instance,  a spin-up electron starts  
at the crest of $z$-spin polarization 
and travels at the Fermi velocity in the direction of the PSH wave vector, its spin precesses precisely 
by an angle of $2\,\pi$ during the time it takes to cover the distance of one PSH wavelength. If the electron propagates off direction, the spin will still match the phase of the PSH everywhere because the larger traveling time to, e.g., the neighboring crest is exactly compensated by a smaller precession frequency. 

One promising progress in this direction is the recent realization of the persistent spin helix in a GaAs/AlGaAs quantum well by Koralek {\it et al.}~\cite{koralek}. They used transient spin grating spectroscopy\cite{cameron} to optically excite a sinusoidal profile of out-of-plane spin polarization with the ``magic'' PSH wave vector. Due to the presence of symmetry breaking effects in a real quantum well, instead of a state with infinite lifetime, two decaying modes were observed. Koralek \etal~named these two modes the symmetry--{\em reduced} and {\em --enhanced} mode---the latter being the PSH.  Although the lifetime of the observed PSH mode is not infinite it is still of the order of $100\, \mathrm{ps}$, exceeding typical transient spin grating lifetimes by two orders of magnitude.  Intriguingly, the temperature dependence of the PSH lifetime displays a maximum close to $100 \, \mathrm{K}$.

In order to improve the lifetimes it is necessary to figure out what the dominant relaxation mechanisms are. The temperature dependence of the PSH lifetime suggests the involvement of electron-electron interactions,~\cite{koralek} which are known to relax spin currents via the spin Coulomb drag effect.\cite{Amico1, flensberg, Amico2, weber}  However, since electron-electron interactions respect the SU(2) symmetry of the PSH state, they cannot be the sole reason for a finite lifetime but in addition a symmetry breaking term must be present.\cite{bernevig1} Here, we consider extrinsic spin orbit interaction~\cite{raimondi} and cubic Dresselhaus spin orbit interaction\cite{stanescu} as a possible source of symmetry breaking as proposed by Koralek {\it et al.}\cite{koralek}.

It is the purpose of this work to develop a theoretical understanding of the PSH lifetime and how the lifetime could be enhanced. In particular we consider the effect of electron-electron interactions in the diffusive D'yakonov-Perel' regime. Regarding symmetry breaking mechanisms, our model (Sec.~\ref{sec:model}) takes into account the effect of extrinsic spin-orbit coupling, which results from the interaction of the conduction electron spins with impurities, as well as the cubic Dresselhaus spin-orbit interaction, which is known to be present in the experimental quantum well to a non-negligible amount.~\cite{koralek} In Sec.~\ref{sec:spindiffeq} we derive a diffusion equation for the spin density in our model system and discuss the contribution of the different symmetry breaking mechanisms. 
In Sec.~\ref{sec:schematic} we present analytical solutions for the simplified situation where only one symmetry breaking mechanism is present.
We propose that a spatially damped spin profile could enhance the lifetime compared to the PSH lifetime. For the parameters of the GaAs/AlGaAs quantum well used by Koralek {\it et al.}~\cite{koralek}  (Sec.~\ref{sec:numbers}) it turns out that electron-electron interactions in combination with cubic Dresselhaus spin-orbit interaction are the key ingredients to understand the temperature dependence of the PSH lifetime. Detailed conclusions and an outlook are given in Sec.~\ref{sec:conclusions}.

\section{Model\label{sec:model}}

In an envelope-function description\cite{winkler} of the conduction band electrons in semiconductor quantum wells, the spin-orbit interaction takes the form of a momentum-dependent, in-plane effective magnetic field. The two dominant contributions to this field are linear in the in-plane momentum: The Rashba field,~\cite{rashba} which has winding number $1$ in momentum space, is caused by structure inversion asymmetry and can be tuned by changing the doping imbalance on both sides of the quantum well. The linear Dresselhaus\cite{dresselhaus} contribution, in contrast, has winding number $-1$ and its physical origin is the bulk inversion asymmetry of the zinc-blende type quantum well material. It is proportional to the kinetic energy of the electron's out-of-plane motion and therefore decreases quadratically with increasing well width. In addition, a small cubic Dresselhaus spin-orbit interaction is present as well.

Thus we write the Hamiltonian for conduction band electrons in the (001) grown quantum well as
\begin{align}
H&~=~H_\textnormal{0}+H_\textnormal{imp}+H_\textnormal{e-e}\label{H}.
\end{align}
The first term represents a two-dimensional electron gas (2DEG) 
with a quadratic dispersion and intrinsic spin-orbit interaction
\begin{align}
H_0&~=~\sum_{s,s'; \vc{k}} \psi_{\vc{k} s'}^\dagger\,\mathcal{H}_{0 s's}\,\psi_{\vc{k} s}\label{general2ndq}
\end{align}
with the $2 \times 2$ matrix in spin space
\begin{align}
\mathcal{H}_0&~=~\epsilon_k+ \vc{b}(\vc{k})\cdot {\bf \vc{\sigma}}.\label{H0}
\end{align}
 The $\psi_{\vc{k} s}^\dagger \left(\psi_{\vc{k} s}\right)$ are creation (annihilation) operators for electrons with momentum $\vc{k}$ and spin projection $s$. Within the standard envelope function approximation\cite{winkler} one finds  $\epsilon_k=\frac{\hbar^2 k^2}{2\, m}$ where $m$ is the effective mass. The vector of Pauli matrices is denoted by $\vc{\sigma}$  and the in-plane spin-orbit field
\begin{align}
\vc{b}(\vc{k})&~=~\vc{b}_R(\vc{k})+\vc{b}_D(\vc{k})\label{bk}
\end{align}
contains Rashba- as well as linear and cubic Dresselhaus spin-orbit interactions\cite{weng} (henceforth $\hbar \equiv 1$), 
\begin{align}
\vc{b}_R(\vc{k})&=
\alpha\,v_F\,
\begin{pmatrix}
k_y\\ -k_x
\end{pmatrix}\label{br},\\
\vc{b}_D(\vc{k})&=v_F\cos 2\phi\left[{\beta'}
\begin{pmatrix}
-k_x\\ k_y
\end{pmatrix}
-\gamma\,\frac{k^3}{4}
\begin{pmatrix}
\cos3\theta\\ \sin3\theta
\end{pmatrix}
 \right]\nonumber\\
&\quad+
v_F\sin 2\phi\left[{\beta'}
\begin{pmatrix}
k_y\\ k_x
\end{pmatrix}
+\gamma\frac{k^3}{4}
\begin{pmatrix}
\sin3\theta\\ -\cos 3\theta
\end{pmatrix}\label{bd}
 \right].
\end{align}
Here, $v_F$ is the Fermi velocity, the angle $\theta$ gives the direction of $\vc{k}$ with respect to the $x$ axis and $\phi$ denotes the angle between the latter and the (100) crystal axis. The strength of the Rashba spin-orbit field is controlled by $\alpha$ and the coefficient for linear Dresselhaus coupling $\beta'$ contains a momentum-dependent renormalization due to the presence of cubic Dresselhaus coupling, 
\begin{align}
 {\beta'}&~=~\beta-\gamma\,{k^2}/{4},\label{betapr}
\end{align}
 where the ``bare'' linear Dresselhaus coefficient $\beta$ is related to the one for cubic Dresselhaus $\gamma$ via $\beta=\gamma \mean{k_z^2}=\gamma\left(\pi/d \right)^2$  ($d$ being the quantum well width). We assume in the following that the spin-orbit interaction is small compared to the Fermi energy $E_F$, i.e., $b_F/E_F\ll 1$, where $b_F\equiv b(k_F)$ with $k_F$ being the Fermi momentum. 

Furthermore, we have included in Eq.~\eqref{H} electron-impurity interactions,
\begin{align}
H_\textnormal{imp}&~=~\frac{1}{V} \sum_{s,s';\vc{k}, \vc{k'}}\psi_{\vc{k'} s'}^\dagger U_{\vc{k'}\vc{k} s's}\, \psi_{\vc{k}s},
\end{align}
(henceforth volume $V\equiv 1$). The impurity potential is a matrix in spin space, 
\begin{align}
\hat{U}_{\vc{k}\vc{k'}}&~=~
V^\textnormal{imp}_{\vc{k}\vc{k'}}\left(\left\{ \vc{R}_i\right\}\right)\left(1+\sigma_z\,\frac{i \lambda_0^2}{4}\,\left[\vc{k}\times\vc{k'}\right]_z\right),\label{U}
\end{align}
where the spin-dependent part arises from extrinsic spin-orbit interaction\cite{raimondi} of the conduction electrons with the impurity potential. In real space, the matrix operator for electron-impurity interactions reads $ \hat{U}(\vc{x})=V^\textnormal{imp}(\vc{x})+i \,{\lambda_0^2}/{4}\left[\vc{\sigma}\times \vc{\nabla}V^\textnormal{imp}(\vc{x}) \right]\cdot  \vc{\nabla}$, with $V^\textnormal{imp}(\vc{x})=\sum_i v(\vc{x}-\vc{R}_i)$, where $v(\vc{x})$ denotes the potential of each single impurity, $\left\{ \vc{R}_i\right\}$ are the impurity positions (eventually to be averaged over) and $\lambda_0$ is a known material parameter ($\lambda_0=4.7\times 10^{-10}\,\textnormal{m}$ for GaAs). Eq.~\eqref{U}, with $V^{\rm imp}_{\vc{k}\vc{k'}}\left(\left\{\vc{R}_j\right\}\right)=\sum_j v(\vc{k'}-\vc{k})\,e^{-i (\vc{k'}-\vc{k})\cdot \vc{R}_j} $, is then  obtained by Fourier transformation. 

Finally, the Hamiltonian \eqref{H} contains electron-electron interactions, 
\begin{align}
H_\textnormal{e-e}&~=~\frac{1}{2} \sum_{\substack{\vc{k_1}\dots \vc{k_{4}}\\ s_1, s_2}}V_{\vc{k_{3}},\vc{k_{4}},\vc{k_{1}},\vc{k_{2}}}\,\psi_{\vc{k_{4}} s_2}^\dagger \psi_{\vc{k_{3}} s_1}^\dagger \psi_{\vc{k_1} s_1}\psi_{\vc{k_2} s_2}
\end{align}
with a Thomas-Fermi screened Coulomb potential of the form $V_{\vc{k_{3}},\vc{k_{4}},\vc{k_{1}},\vc{k_{2}}}\approx \frac{v(|\vc{k_{3}}-\vc{k_{1}}|)}{\epsilon(|\vc{k_{3}}-\vc{k_{1}}|)}$ where $v(q)=\frac{\hbar^2 2\,\pi}{m\, q\, a^*}$ and $\epsilon(q)\approx1+\frac{2}{q\, a^*}$ with $a^*=\frac{\hbar^2 4 \,\pi\, \epsilon_0\,\epsilon_r }{m\, e^2}$ being the effective Bohr radius. For the GaAs dielectric constant we take a standard value, $\epsilon_r=12.9$.

\section{Spin diffusion equations\label{sec:spindiffeq}}

\subsection{Semiclassical kinetic equations}

Our goal is to describe the dynamics of the spin density in real space. Using the nonequilibrium statistical operator method\cite{AltDerivations} (see Ref.~\onlinecite{zubarev}) we derive kinetic equations for the charge and spin components of the Wigner-transformed density matrix
\begin{align}
\hat{\rho}_\vc{k}(\vc{x},t)&~=~n_\vc{k}(\vc{x},t)+\vc{s}_\vc{k}(\vc{x},t)\cdot\vc{\sigma},
\end{align}
where
\begin{align}
\rho_{\vc{k};s s'}(\vc{x},t)&=\int d \vc{r}\, e^{i \vc{k}\cdot \vc{r}}\mean{\psi^\dagger_{s'}(\vc{x}-\vc{r}/2,t)\,\psi_{s}(\vc{x}+\vc{r}/2,t)} .
\end{align} 

If we restrict our calculation to the zeroth order in ${b}/E_F$ and furthermore neglect terms that are nonlinear in the spin density $\vc{s}_\vc{k}(\vc{x},t)$,\cite{HartreeFock} the equations for charge and spin  read 
\begin{align}
\partial_t\,n_\vc{k}+\vc{v}\cdot\partial_{\vc{x}}\,n_\vc{k}&~=~{\mathcal{J}}^\textnormal{imp}_\vc{k}+{\mathcal{J}}^\textnormal{e-e}_\vc{k},\label{charge}\\
2\,\vc{s}_\vc{k}\times\vc{b}(\vc{k})+\partial_t\,\vc{s}_\vc{k}+\vc{v}\cdot \partial_{\vc{x}}\,\vc{s}_\vc{k}&~=~\vc{\mathcal{J}}^\textnormal{imp}_\vc{k}+\vc{\mathcal{J}}^\textnormal{e-e}_\vc{k}\label{spin}
\end{align}
with $v_i=k_i/m$, where the index $i=x,y$ labels the in-plane spatial directions. Note that spin and charge equations decouple in this approximation because the gradient terms containing $\partial_{k_i}\vc{b}({\vc{k}})$, which would couple the spin and charge equations, are of higher order in ${b}/E_F$. Moreover, in the diffusive limit  $b_F\,\tau\ll 1$ (where $\tau$ is the momentum relaxation time), they would yield terms of higher order in this small parameter $b_F\,\tau$.\cite{stanescu, burkov}
On the right-hand side of Eqs.~\eqref{charge}-\eqref{spin}, we have the collision integrals for impurity scattering,
\begin{align}
{\mathcal{{J}}}^\textnormal{imp}_\vc{k}
&=-\sum_\vc{k'}\,W_{\vc{k}\vc{k'}}\,\delta(\Delta \epsilon)\,\Delta n\left\{1+\frac{\lambda_0^4}{16}\left[(\vc{k}\times\vc{k'})_z\right]^2\right\},\\
\vc{\mathcal{{J}}}^\textnormal{imp}_\vc{k}
&=-\sum_\vc{k'}\,W_{\vc{k}\vc{k'}}\,\delta(\Delta \epsilon)\left\{\Delta\vc{s}+\frac{ \lambda_0^2}{2}\left[\vc{k}\times\vc{k'}\right]_z
\begin{pmatrix}
 -s_y'\\
s_x'\\
0
\end{pmatrix}\right.\nonumber\\
&\left.\qquad\qquad\quad\quad
+\frac{\lambda_0^4}{16}\left[\vc{k}\times\vc{k'}\right]_z^2
\begin{pmatrix}
 s_x+s_x'\\
s_y+s_y'\\
s_z-s_z'
\end{pmatrix}
 \right\},\label{Jimpspin}
\end{align}
with the transition rate $W_{\vc{k}\vc{k'}}= {2\,\pi}\,n_i \,|v\left(\vc{k'}-\vc{k}\right)|^2$, where $n_i$ is the impurity concentration, $\Delta \epsilon\equiv \epsilon_k-\epsilon_{k'}$, $ \Delta n \equiv n_k - n_{k'}$ and $\,\Delta \vc{s} \equiv \vc{s}_k- \vc{s}_{k'}$, as well as electron-electron  
scattering,
\begin{align}
\mathcal{J}^\textnormal{e-e}_\vc{k_1}
&=~2\,\pi\sum_{2,3,4}\left(2 |V_{1 2 3 4}|^2- V_{1 2 3 4}V_{1 2 4 3}\right)\delta(\Delta  \tilde{\epsilon})\nonumber\\
&~\quad \left[(1-n_1)(1-n_2)\,n_3\, n_4-\left(1\leftrightarrow 3,~2  \leftrightarrow 4\right) \right],\\
\vc{\mathcal{J}}^\textnormal{e-e}_\vc{k_1}\nonumber
&=~2\,\pi\sum_{2,3,4}\,\delta(\Delta \tilde{\epsilon})\,\label{vectorJee}
\left\{(1-n_1)(1-n_2)\,n_3\,n_4\right. \nonumber\\
&~\quad  \left[ 2 |V_{1 2 3 4}|^2 \left(\frac{\vc{s}_3}{n_3}-\frac{\vc{s}_1}{1-n_1} \right)\right.\nonumber\\
&\quad\quad \left. - V_{1 2 3 4}V_{1 2 4 3}\left(\frac{\vc{s}_3}{n_3}+\frac{\vc{s}_4}{n_4}-\frac{\vc{s}_1}{1-n_1}-\frac{\vc{s}_2}{1-n_2}\right)\right]\nonumber\\
&\quad~ - \left(1\leftrightarrow 3,~2  \leftrightarrow 4\right)\left. \right\}.
\end{align}
Here, we abbreviated $j\equiv \vc{k_j}~$(where $j=1,2,3,4 $ labels initial and final states of the two collision partners) and $\Delta \tilde{\epsilon} \equiv \epsilon_\vc{k_1}+\epsilon_\vc{k_2}-\epsilon_{\vc{k_3}}-\epsilon_{\vc{k_4}}$. 

In our approximation the charge kinetic equation \eqref{charge} decouples from the spin kinetic equation \eqref{spin} and is independently solved by the Fermi-Dirac distribution  $n_\vc{k}(\vc{x},t)=f(\epsilon_k)=\left[1+e^{(\epsilon_k-E_F)/k_B T}\right]^{-1}$, where $k_B$ is the Boltzmann constant and $T$ the temperature. Since we are not interested in charge transport or local charge excitations, we assume that the charge distribution is given by this spatially uniform solution. In the next subsection we use the spin kinetic equation~\eqref{spin} to derive a drift-diffusion equation for the real space spin density, cf.~Refs.~\onlinecite{burkov, mishchenko, stanescu, weng}.

\subsection{Spin diffusion equations in the D'yakonov-Perel' regime}

In the following, we consider the D'yakonov-Perel'\cite{dyakonov} regime of strong scattering and/or weak spin-orbit interaction, $b_F\,\tau\ll 1$. During the time interval $\tau$ between two collisions which alter the momentum of an electron---and thereby $\vc{b}(\vc{k})$---its spin precesses around the spin-orbit field only by the small angle $b_F\,\tau$. This results in a random walk behavior of the spin.\cite{yang} In contrast to the weak scattering limit $b_F\,\tau\gg 1$, the spin polarization is actually stabilized by scattering in the strong scattering regime $b_F\,\tau\ll 1$: the stronger the scattering, the slower the D'yakonov-Perel' spin relaxation---a phenomenon often referred to as ``motional narrowing'' in analogy to the reduction of linewidths in NMR spectroscopy due to disorder in the local magnetic fields. 

In the spirit of the derivation by D'yakonov and Perel'\cite{dyakonov} we will exploit the separation of the timescales that govern the evolution of isotropic (in momentum space) and anisotropic  parts of the spin distribution function. 
Since we deal with a spatially inhomogeneous spin density we also have to assume that the timescale connected to the gradient term in Eq.~\eqref{spin} is large as compared to the transport time, i.e.~$v_F\,q\,\tau\ll1$, where $q$ is a typical wave vector of the Fourier transformed spin density. Thus when speaking of ``orders in $b_F\,\tau$'' in the following, we actually have in mind ``orders in $\max\{b_F\,\tau,\,v_F\,q\,\tau\}$''. 

In order to solve the spin kinetic equation~\eqref{spin} we split off an isotropic component $\vc{S}(\vc{x},t)$ from the spin density $\vc{s}_\vc{k}$ and expand the remaining anisotropic component in winding numbers and powers of momentum $k$, 
\begin{align}
 \vc{s}_\vc{k}&~=~-\frac{2\,\pi}{m}\,f'(\epsilon_k)\,\vc{S}+\vc{s}_{\vc{k},1}+\vc{\tilde{s}}_{\vc{k},1}+\vc{s}_{\vc{k},3},\label{ansatz}
\end{align}
with
\begin{align}
 \vc{s}_{\vc{k},1}&~=~f'(\epsilon_k)\,\frac{k}{m}  \sum_{n=\pm 1} \vc{\delta k_{n}}(\vc{x},t)\,e^{i \,n\, \theta},\label{sk1}\\
 \vc{\tilde{s}}_{\vc{k},1}&~=~f'(\epsilon_k)\,\frac{k^3}{k_F^2\,m}  \sum_{n=\pm 1} \vc{\delta \tilde{k}_{n}}(\vc{x},t)\,e^{i \,n\, \theta},\label{sk1tilde}\\
 \vc{s}_{\vc{k},3}&~=~f'(\epsilon_k)\,\frac{k^3}{k_F^2\,m}  \sum_{n=\pm 3} \vc{\delta k_{n}}(\vc{x},t)\,e^{i \,n\, \theta}.\label{sk3}
\end{align}
The anisotropic components of the distribution function arise due to the gradient term in the Boltzmann equation and the precession around the spin-orbit field. Since the spin-orbit fields~\eqref{br}, \eqref{bd} contain terms with winding numbers $\pm 1$ and $\pm 3$ only these winding numbers have to be considered for the anisotropic part of the spin density to lowest order in $b_F \tau$. Furthermore, one can show that the spin density contains only the same powers of $k$ as the corresponding driving terms in Hamiltonian~\eqref{H0}. Thus we consider a $k$- and a $k^3$-term in the ansatz for the winding number $\pm1$-terms of the spin density~\eqref{sk1} and \eqref{sk1tilde}, because the winding number $\pm 1$-terms of the kinetic equation~\eqref{spin} are the gradient term, the linear Rashba and Dresselhaus spin-orbit fields as well as the renormalization of the linear Dresselhaus term due to cubic Dresselhaus spin-orbit interaction. The winding number $\pm 3$-component of the spin density~\eqref{sk3}, on the other hand, contains only a $k^3$-term because only the cubic Dresselhaus spin-orbit field contributes to winding number $\pm 3$ in the kinetic equation~\eqref{spin}.

In the following we consider point-like impurities, i.e., isotropic scattering with $\tau^{-1} = m\, n_i\,v(0)^2$. Furthermore we assume low temperature $T\ll T_F\equiv E_F/k_B$ and perform a Sommerfeld expansion up to order $(T/T_F)^2$ in all momentum integrations. 
 In this procedure we encounter integrals of the form ($n=2,3,4,6,8$)
 \begin{align}
  \int_0^\infty d \epsilon_k\,f'(\epsilon_k)\,k^n &~=~-k_F^n\,z_n(T)
 \end{align}
 with $z_2=1$ and the Sommerfeld functions (see Appendix~\ref{app:sommerfeld})
\begin{align}
 z_3&~=~1+\frac{\pi^2}{8}\,\frac{T^2}{T_F^2}+\mathcal{O}\left(\frac{T^4}{T_F^4}\right),\label{z3}\\
  z_4&~=~1+\frac{\pi^2}{3}\,\frac{T^2}{T_F^2},\\
 z_6&~=~1+\pi^2\,\frac{T^2}{T_F^2},\\
z_8&~=~1+2\,\pi^2\,\frac{T^2}{T_F^2}+\mathcal{O}\left(\frac{T^4}{T_F^4}\right).\label{z8}
\end{align}

With the goal of obtaining diffusion equations for the real space spin density we start by momentum integration of the kinetic equation, $\frac{1}{(2\pi)^2}\int d\,\vc{k}\left[\mathrm{Eq}.~\eqref{spin}\right]$, using the ansatz \eqref{ansatz}. This  yields the {\it isotropic} equation for the isotropic component of the spin density 
\begin{widetext}
\begin{align}
 \partial_t\,S_x &~=~\frac{k_F^2}{2\, \pi} \left\{\frac{1}{2m} \left(\partial_x \delta \hat k_{c,x} +\partial_y \delta \hat k_{s,x}\right)+\alpha v_F\delta \hat k_{c,z}
- \beta v_F \Big{(} \sin 2\phi \delta \bar k_{c,z} + \cos 2\phi \delta \bar k_{s,z} \Big{)}\right\}-z_4\,\gamma_\mathrm{ey}\,S_x,\label{Sxextr}\\
\partial_t\,S_y &~=~\frac{k_F^2}{2\, \pi}\left\{\frac{1}{2m}\left(\partial_x\delta \hat k_{c,y}+\partial_y\delta \hat k_{s,y}\right)+\alpha v_F \delta \hat k_{s,z}+\beta v_F \left(\sin 2\phi  \delta \overline{\bar k}_{s,z}  -\cos 2\phi  \delta \overline{\bar k}_{c,z}\right)\right\}-z_4\,\gamma_\mathrm{ey}\,S_y,\label{Syextr}\\
\partial_t\,S_z &~=~\frac{k_F^2}{2\, \pi}\left\{\frac{1}{2m}\left(\partial_x \delta \hat k_{c,z}+\partial_y \delta \hat k_{s,z}\right)-\alpha v_F (\delta \hat k_{c,x}+\delta \hat k_{s,y})
+\beta v_F \left[ \sin 2\phi \left(\delta \bar k_{c,x}-\delta \overline{\bar k}_{s,y}\right)+\cos 2\phi \left( \delta \overline{\bar k}_{c,y} +\delta \bar k_{s,x}\right)  \right]\right\}\label{Szextr}
\end{align}
\end{widetext}
with
\begin{align}
\vc{\delta \hat k}_{c(s)} &~=~ \vc{\delta k}_{c(s)} + z_4 \vc{\delta \tilde k}_{c(s)}, \\
\vc{\delta \bar k}_{c(s)} &~=~\vc{\delta \hat k}_{c(s)} \!-\! \zeta (z_4 \vc{\delta k}_{c(s)} + z_6 \vc{\delta \tilde k}_{c(s)}+ z_6 \vc{\delta k}_{c3(s3)}) ,\nonumber \\
\vc{\delta \overline{\bar k}}_{c(s)} &~=~\vc{\delta \hat k}_{c(s)} \!-\! \zeta (z_4 \vc{\delta k}_{c(s)} + z_6 \delta \vc{\tilde k}_{c(s)}- z_6 \vc{\delta k}_{c3(s3)}), \nonumber
\end{align}
where 
\begin{align}
 \zeta&~=~\frac{\gamma\,k_F^2}{4 \,\beta}\label{zeta}
\end{align}
represents the ratio of cubic and linear Dresselhaus coupling strengths and
\begin{equation}
\begin{tabular}{ll}
$\vc{\delta k}_{c( c 3)}~=~2\, \textnormal{Re }\vc{\delta k}_{1(3)},\,\,$ & $\vc{\delta \tilde{k}}_{c}~=~2\, \textnormal{Re }\vc{\delta \tilde{k}}_{1},$\\
$\vc{\delta k}_{s( s 3)} ~=~ -2\, \textnormal{Im }\vc{\delta k}_{1(3)},\,\,$ & $\vc{\delta \tilde{k}}_{s}~=~-2\, \textnormal{Im }\vc{\delta \tilde{k}}_{1}$.
\end{tabular}
\end{equation}

Eqs.\eqref{Sxextr}-\eqref{Szextr} can be seen as continuity equations for the spin density where the anisotropic components $\vc{\delta k}_{c(s)}$,  $\vc{\delta k}_{c(s)3}$ and  $\vc{\delta \tilde k}_{c(s)}$ play the role of (generalized) spin currents. The impurity collision integral~\eqref{Jimpspin} contains a spin-dependent part due to extrinsic spin-orbit interaction, which acts as a drain for in-plane spin-polarization with the Elliot-Yafet relaxation rate\cite{raimondi}
\begin{align}
\gamma_{\textnormal{ey}}&~=~\left(\frac{\lambda_0\, k_F}{2}\right)^4\frac{1}{\tau}\label{Gammaey}.
\end{align}
 This relaxation mechanism can be understood as the net effect of the electron spins precessing by a small angle around the extrinsic spin-orbit field {\it during} the collision with an impurity. Since this field is perpendicular to the electronic motion, i.e., it points in $z$-direction, the $z$ component of the isotropic spin density is unaffected by the Elliot-Yafet mechanism. 

The anisotropic components $\vc{\delta k}_{c(s)}$, $\vc{\delta \tilde k}_{c(s)}$  and $\vc{\delta k}_{c3(s3)}$ can in turn be expressed in terms of the isotropic spin density $S_i$ by integrating the kinetic equation~\eqref{spin} times velocity, where, this time, we omit the time derivative. 
The justification for doing so is that, in order to capture the slow precession-diffusion dynamics of the real space density, we can interpret the time derivative as a coarse-grained one, i.e.~$\partial_t \,\vc{S}\rightarrow \Delta \vc{S}/\Delta t$ with $\Delta t\approx b_F^{-1} \gg \tau$.  Then the fast relaxation of the anisotropic components into the  steady state at the beginning of each time interval $\Delta t$ contributes only in higher order in $b_F\,\tau$ to the average over $\Delta t$. Thus, to leading order, it is sufficient to find the  (quasi-)equilibrium solutions for the anisotropic coefficients. Another way of seeing this is in analogy with the Born-Oppenheimer approximation: similarly to  the fast moving electrons in a molecule, which almost instantaneously find their equilibrium positions with respect to the slowly vibrating nuclei, the  anisotropic parts of the spin distribution quickly adjust to the momentary isotropic spin density. The backaction of the anisotropic parts on the isotropic spin density is then well described using their steady state solution.  

By integrating $\frac{1}{(2\pi)^2}\int d\,\vc{k}\,v_{x(y)}\left[\mathrm{Eq}.~\eqref{spin}\right]$, equating terms of the same order in $k$ and solving for the coefficients, we obtain the following anisotropic equations:
\begin{widetext}
\begin{align}
\delta k_{c,x}&~=~4\pi \left[\alpha v_F  (1+z_4 \gamma_\mathrm{sw} \tau_1) -  \beta v_F \sin 2\phi(1-z_4 \gamma_\mathrm{sw} \tau_1) \right] \tau_1 S_z+ \frac{2\pi}{m} \tau_1 \left(\partial_x S_x + z_4 \gamma_\mathrm{sw} \tau_1 \partial_y S_y \right), \label{kcx}\\
\delta k_{c,y}&~=~-4\,\pi \beta v_F \tau_1 \cos 2\phi\,\left(1-z_4 \gamma_\mathrm{sw} \tau_1\right)\, S_z+\frac{2 \,\pi}{m} \tau_1\left(\partial_x \,S_y-z_4 \gamma_\mathrm{sw} \tau_1\partial_y \,S_x\right),\label{5nnew}\\
\delta k_{c,z}&~=~ 4\,\pi \left(-\alpha v_F+\beta v_F \sin 2\phi\right) \tau_1 S_x+4\,\pi \beta v_F\,\tau_1\cos 2\phi\,S_y  +\frac{2\,\pi}{m} \tau_1 \partial_x \,S_z\label{6new},\\
\delta k_{s,x}&~=~-4\,\pi \beta v_F \tau_1 \cos 2\phi\,\left(1-z_4 \gamma_\mathrm{sw} \tau_1\right)\, S_z +\frac{2\,\pi}{m}\tau_1\left(\partial_y \,S_x-z_4 \gamma_\mathrm{sw} \tau_1\partial_x \,S_y\right)\label{7nnew},\\
\delta k_{s,y}&~=~4\pi \left[\alpha v_F  (1+z_4 \gamma_\mathrm{sw} \tau_1) + \beta v_F \sin 2\phi(1-z_4 \gamma_\mathrm{sw} \tau_1) \right] \tau_1 S_z+\frac{2\,\pi}{m}\tau_1\left(\partial_y \,S_y+z_4 \gamma_\mathrm{sw} \tau_1\partial_x \,S_x\right),\label{deltaksy}\\
\delta k_{s,z}&~=~4\,\pi\beta v_F\,\tau_1 \cos 2\phi\,S_x-4\,\pi \left[\alpha v_F+\beta v_F\sin2\phi\right]\tau_1 S_y  +\frac{2\,\pi}{m}\tau_1 \partial_y \,S_z\label{9nnew},\\
\delta \tilde{k}_{c,x}&~=-\delta \tilde{k}_{s,y}=
~4\,\pi\beta v_F \zeta \sin 2\phi \tilde \tau_1(1-\frac{z_6}{z_4} \gamma_\mathrm{sw} \tilde \tau_1) S_z,\\
\delta \tilde{k}_{c,y}&~=\delta \tilde{k}_{s,x}=
~ 4\,\pi\beta v_F \zeta \cos 2\phi \tilde \tau_1 (1-\frac{z_6}{z_4} \gamma_\mathrm{sw} \tilde \tau_1) S_z,\label{deltaktildecy}\\
\delta \tilde{k}_{c,z}&~=~ -4\,\pi\,\beta v_F \zeta \tilde{\tau}_1 (\sin2\phi\, S_x+\cos2\phi\, S_y),\\
\delta \tilde{k}_{s,z}&~=~ -4\,\pi\,\beta v_F \zeta \tilde{\tau}_1 (\cos 2\phi\, S_x-\sin2\phi\, S_y)\label{deltaktildesz}.
\end{align}
\end{widetext}
The spin densities $S_i$ act as sinks and sources in the equations for the anisotropic coefficients ${\delta k}_{\pm 1,\pm 3,i},{\delta\tilde{k}}_{\pm 1,i}$. Since the spin densities $S_i$ are determined by the initial conditions at $t=0$, they are of zeroth order in $b_F\,\tau$, whereas the anisotropic coefficients ${\delta k}_{\pm 1,\pm 3,i},{\delta\tilde{k}}_{\pm 1,i}$ are already first order in $b_F\,\tau$. Had we included parts with higher winding numbers $\pm2, \pm4, \pm5, \dots$ in our ansatz, these would have been generated only indirectly via the ${\delta k}_{\pm 1,\pm 3,i},{\delta\tilde{k}}_{\pm 1,i}$ (all of which are already of first order in $b_F\,\tau$) and would therefore be of even higher order in $b_F\,\tau$.

In Eqs.~\eqref{kcx}-\eqref{deltaktildesz} we have defined the rate of ``swapping of the spin currents''\cite{lifshits} as
\begin{align}
\gamma_\mathrm{sw}&~=~
\left(\frac{\lambda_0\, k_F}{2}\right)^2\frac{1}{\tau}\label{Gammasw},
\end{align}
which is due to extrinsic spin-orbit interaction like the Elliot-Yafet rate $ \gamma_\mathrm{ey}$ (Eq.~\eqref{Gammaey}), but lower order in $\lambda_0$. It leads to a ``swapping of spin currents'' because a finite $\gamma_\mathrm{sw}$ generates, e.g., a $S_y$ spin current in response to a gradient of the $S_x$ spin density in $x$ direction (see Eq.~\eqref{deltaksy}). Eqs.~\eqref{kcx}-\eqref{deltaktildesz} are valid to linear order in $\tau\,\gamma_\mathrm{sw}\ll 1$. 

Since the anisotropic components $\vc{\delta k}_{\pm1}$ and  $\vc{\delta \tilde k}_{\pm1}$ are related to (generalized) spin currents, the anisotropic  equations \eqref{kcx}-\eqref{deltaktildesz} express generalized Ohm's laws. The effective relaxation times for the anisotropic parts of the spin distribution function are obtained as the inverse sum of the collision integrals for normal impurity scattering, spin-dependent impurity scattering and electron-electron scattering, 
\begin{align}
 \tau_1&~=~\left( \frac{1}{\tau}+\gamma_\mathrm{ey}\,{z_6}+\frac{1}{{\tau}_{\textnormal{e-e},1}}\right)^{-1},\label{tau1}\\
\tilde{\tau}_1&~=~\left( \frac{1}{\tau}+\gamma_\mathrm{ey}\,\frac{z_8}{z_4}+\frac{1}{z_4\,\tilde{\tau}_{\textnormal{e-e},1}}\right)^{-1}.\label{tau1tilde}
\end{align}
Here, the temperature-dependent rates $\tau_{\textnormal{e-e},1}^{-1},\,\tilde{\tau}_{\textnormal{e-e},1}^{-1}$ account for the decay of the respective component ($\vc{s}_{\vc{k},1}$ or $\vc{\tilde{s}}_{\vc{k},1}$) of the spin distribution due to two-particle Coulomb scattering. The rate at which winding-number-$\pm 1$ and linear-in-$k$ components of the spin distribution relax due to electron-electron interaction is  
\begin{align}
{\tau}_{\textnormal{e-e},1}^{-1}
 &=-\frac{1}{k_B\,T\,k_F \,m\,(2 \pi)^4} \iiint {d \vc{k_1}d \vc{k_2}d \vc{k_3}}\,\delta(\Delta\tilde{\epsilon})\,k_1\nonumber\\
 &\qquad  \left[1-f(\epsilon_{k_3})\right]\left[1-f(\epsilon_{\vc{k_1}+\vc{k_2}-\vc{k_3}})\right]f(\epsilon_{k_1})f(\epsilon_{{k_2}})\nonumber\\ &\qquad \left\{2|V(|\vc{k_1}-\vc{k_3}|)|^2    \,\left[\cos  (\theta_3-\theta_1)\,k_3-k_1 \right]      \right.\nonumber\\
&\quad~\left.+V(|\vc{k_1}-\vc{k_3}|)V(|\vc{k_2}-\vc{k_3}|) \right.\nonumber\\
&\quad\left. ~~\left[k_1+\cos  (\theta_2-\theta_1)\,k_2-\cos  (\theta_3-\theta_1)\,k_3 \right.\right.\nonumber\\
&\quad\left.\left. ~-\cos 3(\theta_{1+2-3}-\theta_1)\,|\vc{k_1}+\vc{k_2}-\vc{k_3}|\right] \right\}.\label{tauee1}
\end{align}
It is related to the spin Coulomb drag conductivity from Refs.~\onlinecite{Amico1,flensberg, Amico2} via the Drude formula. The analogous expression for the winding-number-$\pm 1$ but cubic-in-$k$ components reads
\begin{align}
\tilde{\tau}_{\textnormal{e-e},1}^{-1}
 &=-\frac{1}{k_B\,T\,k_F^4 m\,(2 \pi)^4} \iiint {d \vc{k_1}d \vc{k_2}d \vc{k_3}} \,\delta(\Delta\tilde{\epsilon})\,k_1\nonumber\\
 &\qquad \left[1-f(\epsilon_{k_3})\right]\left[1-f(\epsilon_{\vc{k_1}+\vc{k_2}-\vc{k_3}})\right]f(\epsilon_{k_1})f(\epsilon_{{k_2}})\nonumber\\ &\qquad \left\{2|V(|\vc{k_1}-\vc{k_3}|)|^2    \,\left[\cos  (\theta_3-\theta_1)\,k_3^3-k_1^3 \right]      \right.\nonumber\\
&\quad~\left.+V(|\vc{k_1}-\vc{k_3}|)V(|\vc{k_2}-\vc{k_3}|) \right.\nonumber\\
&\quad\left. ~~\left[k_1^3+\cos  (\theta_2-\theta_1)\,k_2^3-\cos  (\theta_3-\theta_1)\,k_3^3 \right.\right.\nonumber\\
&\quad\left.\left. ~-\cos 3(\theta_{1+2-3}-\theta_1)\,|\vc{k_1}+\vc{k_2}-\vc{k_3}|^3\right] \right\}.\label{tauee1tilde}
\end{align}

To find the anisotropic equations for $\vc{\delta k}_{\pm3}$ we follow a similar procedure as before and integrate $\frac{1}{(2\pi)^2}\int d\,\vc{k}\,e^{\pm i 3 \theta}\left[\mathrm{Eq}.~\eqref{spin}\right]$, which results in
\begin{align}
\vc{\delta k}_{c3}&~{=}~{\gamma\, v_F \,k_F^2\,\pi\, {\tau_3}}
\begin{pmatrix}
  \sin 2 \phi\, S_z\\
-\cos 2 \phi \, S_z\\
\cos 2 \phi \, S_y- \sin 2 \phi\, S_x
\end{pmatrix},\label{kc3}\\
 \vc{\delta k}_{s3}&~{=}~{\gamma\, v_F \,k_F^2\,\pi\, {\tau_3}}
\begin{pmatrix}
 \cos 2 \phi \, S_z\\
\sin 2 \phi \, S_z\\
-\sin 2 \phi \, S_y- \cos 2 \phi\, S_x
\end{pmatrix}.\label{ks3}
\end{align}
with
\begin{align}
{\tau}_3 &~{=}~\left( \frac{1}{\tau}+\gamma_\mathrm{ey}\,\frac{z_8}{z_3}+\frac{1}{z_3\,{\tau}_{\textnormal{e-e},3}}\right)^{-1}.\label{tau3}
\end{align}
The electron-electron scattering rate that enters the effective relaxation time \eqref{tau3} for the winding-number-$\pm 3$ parts of the spin distribution is given by 
\begin{align}
\tau_{\textnormal{e-e},3}^{-1} &=- \frac{1}{k_B\,T\,k_F^3 m\,(2 \pi)^4}\iiint {d \vc{k_1}d \vc{k_2}d \vc{k_3}} \,\delta(\Delta\tilde{\epsilon})\nonumber\\
 &\qquad \left[1-f(\epsilon_{k_3})\right]\left[1-f(\epsilon_{\vc{k_1}+\vc{k_2}-\vc{k_3}})\right]f(\epsilon_{k_1})\,f(\epsilon_{{k_2}})\nonumber\\ &\qquad \left\{2|V(|\vc{k_1}-\vc{k_3}|)|^2    \,\left[\cos 3 (\theta_3-\theta_1)\,k_3^3-k_1^3 \right]      \right.\nonumber\\
&\quad~\left.+V(|\vc{k_1}-\vc{k_3}|)V(|\vc{k_2}-\vc{k_3}|) \right.\nonumber\\
&\quad\left. ~~\left[k_1^3+\cos 3 (\theta_2-\theta_1)\,k_2^3-\cos 3 (\theta_3-\theta_1)\,k_3^3 \right.\right.\nonumber\\
&\quad\left.\left. ~-\cos 3 (\theta_{1+2-3}-\theta_1)\,|\vc{k_1}+\vc{k_2}-\vc{k_3}|^3\right] \right\}.\label{tauee3}
\end{align}

Finally we insert the steady-state solutions for the anisotropic coefficients~\eqref{kcx}-\eqref{deltaktildesz} and \eqref{kc3}-\eqref{ks3} into the isotropic equations~\eqref{Sxextr}-\eqref{Szextr} and obtain a closed set of coupled diffusion equations for the three spatial components of the spin density,
\begin{widetext}
\begin{align}
\partial_t\,\vc{S}
&=
 \begin{pmatrix}  {D}\,\grad^2-{\Gamma}_x-{\gamma}_{\textnormal{cd}}\,z_6-\gamma_\mathrm{ey}\,z_4&~&{L}&~&{K}_{xz}\,\partial_x-
{M}\,\partial_y\\
{L}&~& {D}\,\grad^2-{\Gamma}_y-{\gamma}_{\textnormal{cd}}\,z_6-\gamma_\mathrm{ey}\,z_4&~&{K}_{yz}\,\partial_y-
{M}\,\partial_x\\
-{K}_{zx}\,\partial_x+{M}_{z}\,\partial_y&~&-{K}_{zy}\,\partial_y+{M}_{z}\,\partial_x&~& {D}\,\grad^2-{\Gamma}_x-{\Gamma}_y-2 \,{\gamma}_{\textnormal{cd}}\,z_6-\Gamma_\mathrm{sw}
 \end{pmatrix}
 \vc{S}.\label{matrixeq}
\end{align}
\end{widetext}
On its diagonal  the matrix operator contains the pure diffusion terms with $ \grad^2=\partial_x^2+\partial_y^2$ and the Elliot-Yafet relaxation rate $\gamma_{\rm ey}$ due to extrinsic spin-orbit interaction. In addition,  it contains the D'yakonov'-Perel' relaxation rates $\Gamma_{x(y)}$ and $\gamma_{\textnormal{cd}}$ which reflect the randomization of the spin orientation due to precession (between the collisions) around the winding-number-$\pm 1$ and winding-number-$\pm 3$ spin-orbit fields, respectively. 
The $S_x$ component is relaxed as a consequence of precession about the $y$ component of the spin-orbit field only, and vice versa. In contrast, the $S_z$ component is relaxed by the precession about the full spin-orbit field. Thus the relaxation rate of $S_z$ due to precession is the sum of the ones for $S_x$ and $S_y$, plus a correction $\Gamma_\mathrm{sw}$ for processes that involves the swapping of the spin currents due to extrinsic spin-orbit interaction. Due to precession there are also off-diagonal rates $L$, which couple the in-plane spin components, as well as several off-diagonal mixed diffusion-precession rates, which are accompanied by partial derivatives.

In terms of the parameters of our model and previously defined quantities, the coefficients in the spin diffusion equation~\eqref{matrixeq} are given by:
\begin{widetext}
\begin{align}
\gamma_{\textnormal{cd}}&~{=}~\frac{1}{8}\,v_F^2\,\gamma^2\,k_F^6 \,{\tau}_3,\label{Gammacd}\\
 \Gamma_{x(y)}(\phi)&~{=}~\frac{1}{4}\,q_0^2\left(D\mp\frac{\beta}{\alpha}\left[2 \,D-\zeta\,z_4\,(D+\tilde{D})\right]\sin2\phi+\frac{\beta^2}{\alpha^2}\left[D-\zeta\,z_4\,(D+\tilde{D})+\zeta^2z_6\,\tilde{D}\right]\right),\label{Gxysw}\\
\Gamma_\mathrm{sw}&~{=}~\frac{1}{2}\,q_0^2\,\gamma_\mathrm{sw}\left[D\,\tau_1\,z_4-\frac{\beta^2}{\alpha^2}\left(D\,\tau_1\,z_4-\zeta\,\tilde{D}\,\tilde{\tau}_1\,z_6-\zeta\,D\,\tau_1\,z_4^2+\zeta^2\,\tilde{D}\,\tilde{\tau}_1\,\frac{z_6^2}{z_4}\right)\right],\\ K_{xz(yz)}(\phi)&~{=}~q_0\left(D\mp\frac{\beta}{\alpha}\left[D-\frac{1}{2}\,\zeta\,z_4\,(D+\tilde{D})\right]\sin2\phi\right)+\frac{1}{2}\,\gamma_\mathrm{sw}\,q_0\left(\tau_1\,D\,z_4\pm\frac{\beta}{\alpha}\left[\tau_1\,D\,z_4-\zeta\,\tilde{\tau}_1\,\tilde{D}\,z_6\right]\sin 2 \phi\right),\label{Kxzsw}\\
K_{zx(zy)}(\phi)&~{=}~q_0\left(D\mp\frac{\beta}{\alpha}\left[D-\frac{1}{2}\,\zeta\,z_4\,(D+\tilde{D})\right]\sin2\phi\right)+\frac{1}{2}\,\gamma_\mathrm{sw}\,\tau_1\,q_0\,D\,z_4\left[1\pm\frac{\beta}{\alpha}\left(1-\zeta\,z_4\right)\sin2\phi\right],\label{Kzxsw}\\
M(\phi)&~{=}~\cos 2\phi\,q_0\,\frac{\beta}{\alpha}\left[D-\frac{1}{2}\,\zeta\,z_4\,(D+\tilde{D})\right]-\frac{1}{2}\,\gamma_\mathrm{sw}\,q_0\,\cos2\phi\,\frac{\beta}{\alpha}\left[\tau_1\,D\,z_4-\zeta\,\tilde{\tau}_1\,\tilde{D}\,z_6\right],\label{Mxysw}\\
M_{z}(\phi)&~{=}~\cos 2\phi\,q_0\,\frac{\beta}{\alpha}\left[D-\frac{1}{2}\,\zeta\,z_4\,(D+\tilde{D})\right]-\frac{1}{2}\,\gamma_\mathrm{sw}\,\tau_1\,q_0\,D\,z_4 \,\cos 2\phi\,\frac{\beta}{\alpha}\left(1-\zeta\,z_4\right),\label{Mzsw}\\
L(\phi)&~{=}~\cos 2\phi\,\frac{1}{2}\,q_0^2\,\frac{\beta}{\alpha}\left[D-\frac{1}{2}\,\zeta\,z_4\,(D+\tilde{D})\right]\label{L}
\end{align}
\end{widetext}
with the PSH wave vector 
\begin{align}
 {q}_0&~=~4\,v_F\,m\,\alpha \label{magicq}
\end{align}
and the effective diffusion constants 
\begin{equation}
 D=\frac{1}{2}\, v_F^2\,\tau_1,\quad\tilde{D}=\frac{1}{2}\, v_F^2\,\tilde{\tau}_1.
\end{equation}
At $T=0$, we have $z_n=1$ and electron-electron interactions are absent, such that $\tilde{D}=D $. Then, if we leave out extrinsic spin-orbit interaction in the spin diffusion equation \eqref{matrixeq}, it agrees with the one 
presented in Ref.~\onlinecite{weng} (except for the sign of $L$).
 If we further omit cubic Dresselhaus spin-orbit interaction in our diffusion equation, it also concurs with the one of Ref.~\onlinecite{bernevig1} 
 provided that the spin-charge coupling is negligible.

\section{Persistent spin helix in the presence of symmetry breaking mechanisms\label{sec:schematic}}


In this section, we use the spin diffusion equation~\eqref{matrixeq} to calculate the lifetime of the persistent spin helix 
in the presence of symmetry breaking mechanisms. We consider 
extrinsic spin-orbit interaction, cubic Dresselhaus spin-orbit interaction or simple spin-flip scattering as possible symmetry breaking mechanisms. In order to allow for simple analytical solutions we discuss each of the symmetry breaking mechanisms separately. In the case of cubic Dresselhaus spin-orbit interaction we neglect at first the renormalization of the linear Dresselhaus spin-orbit interaction (see~Eq.~\eqref{betapr}). This is formally achieved by setting $\zeta = 0$ in Eqs.~\eqref{Gxysw}-\eqref{L} while keeping the $ {\gamma}_\textnormal{cd}$ term in Eq.~\eqref{matrixeq}. However, we will include the renormalization of the linear Dresselhaus spin-orbit interaction when we discuss a possible stationary solution and when we compare to the experimental results in a GaAs/AlGaAs quantum well  in Sec.~\ref{sec:numbers}.

We choose our coordinate system such that the $x$ axis points into the (110)-crystal direction, corresponding to $\phi=\frac{\pi}{4}$ in Eqs.~\eqref{Gxysw}-\eqref{L}. Considering an initial spin polarization, which is uniform in $x$-direction, then due to $L(\frac{\pi}{4})=M(\frac{\pi}{4})=0$ the $S_x$ component decouples from the $S_y$ and $S_z$ components and we can set $S_x=0$. For $\alpha=\beta$ Eq.~\eqref{matrixeq} reduces for  the remaining $S_y$ and $S_z$ components to 
\begin{align}
\partial_t\,\vc{S}&~=~
\begin{pmatrix}
{D}\,\partial_y^2-q_0^2\,D-X &2\,q_0\,D\,\partial_y\\
-2\,q_0\,D\,\partial_y& {D}\,\partial_y^2-q_0^2\,D-N\,X 
\end{pmatrix}\,\vc{S},\label{2by2eqatSP}
\end{align}
where the relaxation rates due to the respective symmetry-breaking  mechanism are represented by $X$ and an integer $N$ according to Table~\ref{tab:XN}.
\begin{table}
\caption{Specification of $X$ and $N$ in Eq.~\eqref{2by2eqatSP} 
\label{tab:XN}}
\vspace{0.2 cm}
\begin{tabular}{c|c|c|c}
 &\,\,\textnormal{simple spin flips}\,\,&\,\textnormal{extr.~spin-orbit int.}&\, \textnormal{ cubic Dress. }\,\\
\hline
$X$\,&$1/\tau_\textnormal{sf}$&$\,\,\, \gamma_\textnormal{ey}$ &$\,\gamma_\textnormal{cd}\,z_6$\\
$N$\,&$1$ & 0 &$2$
\end{tabular}
\end{table}

For the SU(2) symmetric situation $X=0$ there exists a steady state solution with wave vector $q_0$. This is the persistent spin helix state. More precisely for an initial spin polarization of the form
\begin{align}
\vc{S}(\vc{x},t=0)&~=~S_0\left(0,\,0,\,\cos q_0 y\right),\label{initial}
\end{align}
similar to the experimental set-up,\cite{koralek} one finds that the time-dependent solution to Eq.~\eqref{2by2eqatSP} is
\begin{align}
 \vc{S}^{X=0}(y,t)&~=~\frac{S_0}{2} 
\begin{pmatrix}
[e^{-4\,q_0^2\,D\,t}-1]\,\sin q_0 y\\
[e^{-4\,q_0^2\,D\,t}+1]\,\cos q_0 y
\end{pmatrix}.\label{symmetric}
\end{align}
For $t\rightarrow \infty$, i.e., in the stationary limit, this reduces to the persistent spin helix state. 

In the presence of  symmetry breaking mechanisms, i.e. for $X \ne 0$, one can still find a steady state solution of the form
\begin{align}
S_y(y)&~=~-\frac{S_0}{2}\,e^{-{y}/{l_\textnormal{X}}}\,C_1\,\sin q_\textnormal{X}y,\label{dampedy}\\
S_z(y)&~=~\frac{S_0}{2}\,e^{-{y}/{l_\textnormal{X}}}\left(C_2\,\sin q_\textnormal{X}y+\cos q_\textnormal{X}y\right).\label{dampedz}
\end{align}
This solution 
is a spatially damped persistent spin helix state with coefficients given by 
\begin{align}
l_{X}^{-1}&~=~\frac{q_0}{2}\,\sqrt{2\,\Xi+(N+1)\,\xi-2},\label{lx}\\
q_{X}&~=~\frac{q_0}{2}\,\sqrt{2\,\Xi-(N+1)\,\xi+2},\label{qx}\\
C_1&~=~\frac{4\,\sqrt{2\,\Xi-(1+N)\,\xi+2}}{\xi^2\,(-8(N^2-1)+(N-1)^3\xi)}\nonumber\\
&\qquad\Big{[}4+(3\,N+1)\,\xi -N\,(N-1)\,\xi^2\nonumber\\
&\qquad\,-(4+(N-1)\,\xi)\,\Xi\,\Big{]},\\
C_2&~=~\frac{8-(N-1)^2\xi^2-4\left(2\,\Xi-(N+1)\,\xi\right)}{(N-1)\,\xi\,\sqrt{8\,(N+1)\xi-(N-1)^2\xi^2}},\label{C2}
\end{align}
where $\xi\equiv X/({q_0^2\,D})$ and $\Xi\equiv\sqrt{(1+\xi)(1+N\,\xi)}$. In the absence of  symmetry breaking mechanisms ($\xi\rightarrow 0$)  the $t\rightarrow \infty$ asymptotics of Eq.~\eqref{symmetric}, i.e., the truely persistent spin helix state, is recovered. The spatially damped persistent spin helix state \eqref{dampedy}-\eqref{dampedz} could in principle be excited with the initial spin polarization profile
\begin{align}
 \vc{S}(\vc{x},t=0)&~=~S_0\,e^{-{y}/{l_\textnormal{X}}}\left(0,\,0,\,\cos q_\textnormal{X} y\right).\label{damped}
\end{align}

Although the spatially damped persistent spin helix is clearly a steady state solution when the symmetry breaking is caused by simple spin flips or extrinsic spin orbit interaction, it is not obvious that this applies also to the case of cubic Dresselhaus spin orbit interaction, since we have neglected the renormalization of the linear Dresselhaus spin orbit interaction ($\zeta \ne 0$), which might lead to a finite lifetime of the spatially damped state. Nevertheless, even when the renormalization of the linear Dresselhaus spin orbit interaction is taken into account one can still find a steady state solution of the form~\eqref{dampedy}-\eqref{C2}  when the ratio of the linear Rashba and Dresselhaus spin orbit interactions is given by
\begin{align}
 \frac{\beta}{\alpha}&~=~\frac{D}{D-\frac{1}{2}\,\zeta\,z_4\,(D+\tilde{D})}.\label{B1}
\end{align} 
Then the spin diffusion equation~\eqref{matrixeq} can still be cast into the form of Eq.~\eqref{2by2eqatSP} when the symmetry breaking rate $X$ is redefined as $\tilde{X}=  X+ q_0^2 D\, F(T)$ with the temperature dependent dimensionless function $F(T)= \frac{1}{4}\Big(\frac{D^2-\zeta z_4\,D\,(D+\tilde{D})+\zeta^2 z_6\, D\,\tilde{D}}{D^2-\zeta z_4 D(D+\tilde{D})+\frac{1}{4}\zeta^2 z_4^2( D+\tilde{D})^2}-1\Big).$
For this symmetry breaking rate $\tilde{X}$ and spin orbit couplings of \eqref{B1} the spatially damped spin profile of the form \eqref{dampedy}-\eqref{C2} is again infinitely long-lived.

This stationary state should in principle be realizable in the GaAs/AlGaAs quantum well used in Ref.~\onlinecite{koralek} because there the ratio of $\beta/\alpha$ almost fulfills relation~\eqref{B1} at a temperature of $T=100$ K. For the parameters of the GaAs/AlGaAs quantum well of Ref.~\onlinecite{koralek} the steady state solution \eqref{dampedy}-\eqref{C2} would be characterized by a wavevector of $q_{\tilde{X}}\approx q_0$ and a damping length of a bit more than a PSH wavelength, $l_{\tilde{X}}\approx 1.06\,\frac{2\,\pi}{q_0}$. Although a spin grating with such a strong spatial damping might be difficult to realize, it should be noted that the required damping length is $ \propto \zeta^{-1}$, so that one can expect much longer damping lengths for thinner quantum well.

We now want to consider the conventional PSH solution. When we stick to an initial spin polarization with the form of a plane wave~\eqref{initial} similar to the experimental set-up~\cite{koralek} the time dependent solution is given by a double exponential decay, 
\begin{align}
 S_y(y,t)&~=~\frac{S_0}{2}\,\sin q_0y\,\frac{4\,q_0^2\,D\left(e^{-\frac{t}{\tau_R}}- e^{-\frac{t}{\tau_E}}\right)}{\sqrt{(4\,q_0^2\,D)^2+(N-1)^2\, X^2}},\label{Sydouble}\\
 S_z(y,t)&~=~\frac{S_0}{2}\,\cos q_0y\left[e^{-\frac{t}{\tau_R}}+ e^{-\frac{t}{\tau_E}}\right.\nonumber\\
&\quad\quad\qquad\left.+\frac{(N-1)\,X\left(e^{-\frac{t}{\tau_R}}- e^{-\frac{t}{\tau_E}}\right)}{\sqrt{(4\,q_0^2\,D)^2+(N-1)^2\, X^2}} \right]\label{Szdouble}
\end{align}
with the symmetry--{\it enhanced} and {\it--reduced} lifetimes
 \begin{align}
\tau_{E(R)}^{-1}&~=~2\,q_0^2\,D+\frac{1}{2}\,(N+1)\,X\nonumber\\
&~\quad\mp\frac{1}{2}\sqrt{(4\,q_0^2\,D)^2+(N-1)^2\, X^2} .\label{tauexact}
\end{align} 
In the absence of any symmetry-breaking relaxation mechanism, i.e., for $X=0$, the proper persistent spin helix state is recovered ($\tau_E=\infty$). Expanding Eq.~\eqref{tauexact} for small $X/(4\,q_0^2\,D)\ll 1$ we obtain
\begin{align}
\tau_{E}&~\approx~     \frac{2}{(N+1)}\,X^{-1}+\left(\frac{N-1}{N+1}\right)^2\frac{1}{4\,q_0^2\,D}       ,\label{tauE}\\
\tau_{R}&~\approx~       \frac{1}{4\,q_0^2\,D} -\frac{(N+1)\,X}{2\,(4\,q_0^2\,D)^2}           \label{tauR}.
\end{align} 
The reduced lifetime $\tau_{R}$ is not very sensitive to details of the symmetry-breaking mechanism as long as it is weak. Correspondingly,  the temperature dependence of the reduced lifetime $\tau_R$ is almost independent of the symmetry breaking mechanism (and is given by the electron-electron relaxation rate $\tau_{\textnormal{e-e},1}^{-1}$ contained in $D$ via $\tau_1$, see~Eq.~\eqref{tau1}). The temperature dependence of the enhanced lifetime $\tau_E$, by contrast, 
depends crucially on the symmetry breaking mechanism under consideration and thus offers a way to discriminate between the different symmetry breaking mechanisms. For small symmetry breaking terms the enhanced lifetime $\tau_E$ is proportional to the respective scattering rate $X^{-1}$. Therefore also the temperature dependence of $\tau_E$ is determined by the respective scattering rate. For simple spin-flip scattering $X=\tau_{\rm sf}^{-1}$ we expect a temperature independent lifetime $\tau_E$ due to constant $\tau_{\rm sf}$. For extrinsic spin-orbit interactions, $X=\gamma_{\rm ey}$, to leading order  in $X/(4\,q_0^2\,D)$ the only temperature dependence comes from the Sommerfeld corrections. Thus $\tau_E$ decreases quadratically with temperature. For cubic Dresselhaus spin-orbit interaction one finds
\begin{align}
\tau_E&~\approx~ \frac{2}{3}\, \gamma_\textnormal{cd}^{-1}z_6^{-1} 
\label{tauEcD}
\end{align}
and therefore $\tau_E$ is 
proportional to 
 $\tau_3^{-1}$ (see Eq.~\eqref{Gammacd}). Since $\tau_3$ decreases with temperature because of enhanced electron-electron scattering $\tau_{\textnormal{e-e},3}^{-1}$ (see Eq.~\eqref{tau3}) the lifetime $\tau_E$ increases initially with temperature due to the motional narrowing effect in the D'yakonov-Perel' regime. The presence of the Sommerfeld function $z_6$, on the other hand, leads to a decrease of $\tau_E$ with increasing temperature. Thus for cubic Dresselhaus spin-orbit interaction we find that the temperature dependence is governed by a competition between increasing and decreasing contributions. We will compare this theoretical interpretation with experimental results for the persistent spin helix in GaAs/AlGaAs quantum wells~\cite{koralek} in the next section.

\section{Persistent spin helix in G\lowercase{a}A\lowercase{s}/A\lowercase{l}G\lowercase{a}A\lowercase{s} quantum wells}
\label{sec:numbers}

\begin{figure}
\psfrag{u}[bl]{{$T$/K}}
\psfrag{v}[bc]{{$\tau^{-1}_{\textnormal{e-e},1(3)},\tilde{\tau}^{-1}_{\textnormal{e-e},1}$/ps}}
\psfrag{x}[bl]{{$T$/K}}
\psfrag{y}[bc]{{$\tau^{-1}_{1(3)},\tilde{\tau}^{-1}_{1}$/ps}}
$\begin{array}[b]{l}

\multicolumn{1}{c}{\mbox{{ \bf{(a)}}}} 	 \\ [-0.2 cm]
\includegraphics[width=0.56\columnwidth,clip=true ]{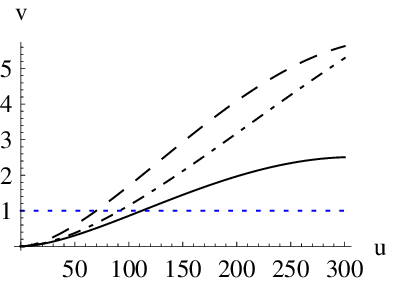}\\
{\vspace{0.2 cm}}\\
\multicolumn{1}{c}{\mbox{{ \bf{(b)}}}} 	 \\ [-0.2 cm]
\includegraphics[width=0.56\columnwidth,clip=true ]{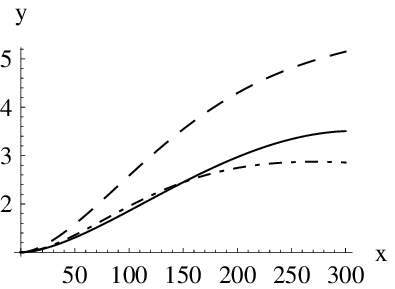}
\\
\end{array}$
\caption{(a) Temperature-dependent relaxation rates due to electron-electron interactions $\tau^{-1}_{\textnormal{e-e},1}$ (solid line), $\tilde{\tau}^{-1}_{\textnormal{e-e},1}$ (dot-dashed line) and $\tau^{-1}_{\textnormal{e-e},3}$ (dashed line), as numerically computed using the experimental parameters of Ref.\onlinecite{koralek}.  In order to continuously interpolate between the data points, we made a fit to the functional form  $A\,T^2+B\,T^2\ln T$, which has been shown to be correct for the spin Coulomb drag conductivity at low temperatures in Ref.~\onlinecite{Amico2}. For comparison we also show (blue dotted line) the inverse transport time $\tau^{-1}$(100 K). 
(b) The resulting effective relaxation rates $\tau^{-1}_{1}$ (solid), $\tau^{-1}_{3}$ (dashed) and $\tilde{\tau}^{-1}_{1}$ (dot-dashed), cf.~Eqs.~\eqref{tau1}-\eqref{tau1tilde} and \eqref{tau3}.}
\label{fig:eerates}
\end{figure}

In order to address the lifetime of the PSH observed experimentally in GaAs/AlGaAs quantum wells\cite{koralek} we consider cubic Dresselhaus alongside with extrinsic spin-orbit interaction as possible symmetry breaking mechanisms. We also include also the renormalization of the linear Dresselhaus coupling constant due to cubic Dresselhaus spin-orbit interaction  ($\zeta \neq 0$ in Eqs.~\eqref{Gxysw}-\eqref{L}). Analogously to the previous section we can set $S_x=0$ and then the spin diffusion equation~\eqref{matrixeq} reduces for the remaining components $S_y$ and $S_z$ to:
\begin{align}
\partial_t\,\vc{S}&~=~
\begin{pmatrix}
{D}\,\partial_y^2-Y &K_{yz}(\pi/4)\,\partial_y\\
-K_{zy}(\pi/4)\,\partial_y& {D}\,\partial_y^2-Z 
\end{pmatrix}\,\vc{S}\label{2by2eq}
\end{align}
with
\begin{align}
Y&~=~\Gamma_y(\pi/4) +\gamma_\textnormal{cd}\,z_6+\gamma_\textnormal{ey}\,z_4,\\
Z&~=~\Gamma_x(\pi/4)+\Gamma_y(\pi/4)+2\,\gamma_\textnormal{cd}\,z_6+\Gamma_\textnormal{sw}.
\end{align}
For an initial spin polarization of the form $\vc{S}(\vc{x},t=0)=S_0\left(0,\,0,\,\cos q_0 y\right)$ the time dependent part of the solution is given by a double exponential decay, i.e., a sum of two exponentially decaying terms with a symmetry enhanced relaxation rate $\tau_E$ and a symmetry reduced relaxation rate $\tau_R$ given by
\begin{align}\label{tauER}
\tau_{E(R)}^{-1}&~=~\frac{1}{2}\,(Y+Z)+q_0^2\,D\\
&\qquad\mp\frac{1}{2}\,\sqrt{(Y-Z)^2+4 \, q_0^2\,K_{yz}(\pi/4)\,K_{zy}(\pi/4)}.\nonumber
\end{align}

In order to compare our theory with the experiment of Ref.~\onlinecite{koralek} we need to calculate the coefficients that occur in Eq.~\eqref{tauER}---in particular the temperature-dependent rates for electron-electron scattering. Fig.~\ref{fig:eerates} (a) shows $\tau^{-1}_{\textnormal{e-e},1}$, $\tilde{\tau}^{-1}_{\textnormal{e-e},1}$ and $\tau^{-1}_{\textnormal{e-e},3}$ , evaluated from Eqs.~\eqref{tauee1}-\eqref{tauee3} by Monte Carlo integration for parameters of Ref.~\onlinecite{koralek}. With  these electron-electron scattering rates  we find for the effective scattering rates in Eqs.~\eqref{tau1}-\eqref{tau1tilde} and \eqref{tau3} the results depicted in Fig.~\ref{fig:eerates}(b).

\begin{figure}[t]
\psfrag{x}[bl]{{$T$/K}}
\psfrag{y}[bc]{{$\tau_{E(R)}$/ps}}
\psfrag{u}[bc]{{$\tau_{E}$/ps}}
$\begin{array}[b]{l}
\multicolumn{1}{c}{\mbox{{ \bf{(a)}}}} 	 \\ [-0.86 cm]
\includegraphics[width=0.66\columnwidth,clip=true ]{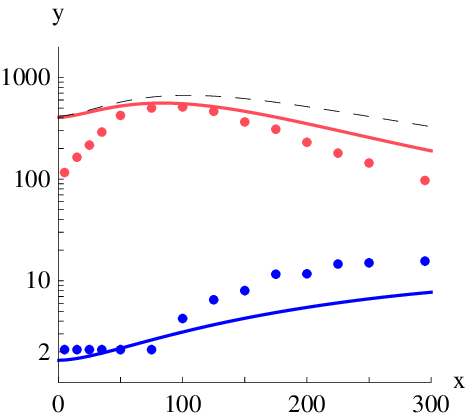}\\
{\vspace{0.4 cm}}\\
\multicolumn{1}{c}{\mbox{{ \bf{(b)}}}} 	 \\ [-0.3 cm]
\hspace{0.2 cm}\includegraphics[width=0.66\columnwidth,clip=true ]{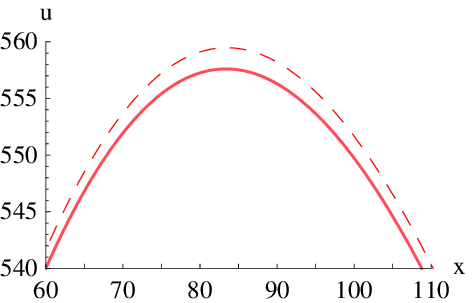}\\
{\vspace{0.4 cm}}\\[0.36 cm]
\multicolumn{1}{c}{\mbox{{ \bf{(c)}}} }	 \\ [-0.86 cm]
\includegraphics[width=0.66\columnwidth,clip=true ]{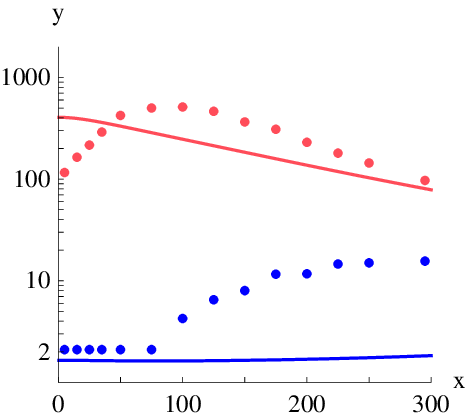}\\
\end{array}$
\caption{(Color online) (a) Temperature-dependent lifetimes of the enhanced (red/grey) and reduced (blue/black) modes. The points are experimental data from Ref.~\onlinecite{koralek}. Solid lines are the respective theoretical curves including extrinsic and cubic Dresselhaus spin-orbit interactions as well as electron-electron interactions; the thin dashed line is the simplified result from Eq.~\eqref{tauEcD}. 
In panel (b) we zoom in on the theoretical curve of panel (a) close to the maximum of $\tau_{E}$. The dashed line is the theoretical curve without extrinsic spin-orbit interaction. Panel (c) depicts the results of a calculation, where we include extrinsic and cubic Dresselhaus spin-orbit interaction but exclude electron-electron interactions. (Also here,  a comparison as in (b) would show that the influence of the extrinsic spin-orbit interactions is marginal.)} 
\label{lifetimeplot}
\end{figure}

In Fig.~\ref{lifetimeplot}, we show the numerical results for the temperature dependence of the symmetry-enhanced and reduced lifetimes $\tau_{E(R)}$ where we use the experimental parameters of Ref.~\onlinecite{koralek}. In particular, we take $\alpha=0.0013$ for the Rashba spin-orbit interaction and $\gamma\,v_F=5.0\textnormal{ eV}\textnormal{ \r{A}}^3$ for the cubic Dresselhaus spin-orbit interaction. We adjust the linear Dresselhaus spin-orbit interaction to $\beta= 1.29 \,\alpha$ in order to maximize $\tau_E$ for $T=75$ K---the temperature at which also in the experiment the spin-orbit interaction was tuned in order to maximize $\tau_E$.

At intermediate temperatures around $100$ K, i.e., in the temperature range where our theory should be most applicable, we find very good agreement between our theory (solid lines) and the experimental lifetimes (dots), see Fig.~\ref{lifetimeplot} (a). We observe a maximum in $\tau_E$ roughly where the experimental points exhibit one. Also the size of $\tau_E$ as well as of $\tau_R$ is very close to the experimental values.
Since the scattering rates due to extrinsic spin-orbit interaction are very small in the GaAs/AlGaAs quantum well under consideration, i.e., $\gamma_\textnormal{ey}/\gamma_\textnormal{cd}\approx 10^{-4}$ and $\tau\,\gamma_\textnormal{sw}\approx 3\times 10^{-3} $, effects of extrinsic spin-orbit interaction turn out to be negligible, see Fig.~\ref{lifetimeplot} (b). A calculation which includes extrinsic spin-orbit interactions and electron-electron interactions but excludes cubic Dresselhaus spin-orbit interaction (not depicted in Fig.~2) would yield enhanced lifetimes that exceed the experimental ones by a factor $\sim 10^3$.

Interestingly, the simple 
result \eqref{tauEcD} 
for the enhanced lifetime, 
where we neglected the renormalization of the linear Dresselhaus spin-orbit interaction due to cubic Dresselhaus spin-orbit interaction, is a fairly good approximation (see dashed curve in Fig.~\ref{lifetimeplot} (a)). Thus the simple interpretation of the temperature dependence of $\tau_E$ can also be extended to the present situation. The formation of the maximum in $\tau_E$ at intermediate temperatures around $100$ K is caused by the competition between two effects: on the one hand $\tau_E$ increases with temperature due to increasing electron-electron scattering, which leads in the presence of symmetry breaking cubic Dresselhaus spin-orbit interaction to the usual motional-narrowing effect in the D'yakonov-Perel' regime. On the other hand the magnitude of Sommerfeld corrections increases with temperature reducing the lifetime $\tau_E$ in two ways: (i) by increasing the effective cubic Dresselhaus scattering rate $\gamma_{\rm cd}\,z_6$ and (ii) by increasing the linear renormalization of the Dresselhaus spin-orbit interaction, which leads to a detuning of the Rashba and the effective linear Dresselhaus spin-orbit interactions.

The important effect of electron-electron interaction for the temperature dependence of the lifetimes $\tau_E$ and $\tau_R$ can also be deduced from Fig.~\ref{lifetimeplot}(c), where we show the lifetimes excluding the effect of electron-electron interactions. Obviously the initial increase of the lifetimes with temperature is absent for both $\tau_E$ and $\tau_R$ in the absence of electron-electron interaction. 

At low temperatures and at high temperatures deviations between our theory and the experimental lifetimes are more pronounced. 
We suppose that  at high temperatures  symmetry breaking mechanisms that are not captured by our model (e.g.~effects involving phonons) could become important. Furthermore, since the Fermi temperature in the GaAs/AlGaAs quantum well under consideration is only $T_F=400$ K we cannot expect our calculation, which is based on a low-order Sommerfeld expansion, to be as accurate in the high temperature range above $200$ K. The disagreement at low temperatures, on the other hand, results most likely from the fact that we do not take into account the temperature dependence of the transport lifetime but rather use the experimental $100$ K-transport lifetime $\tau(100\textnormal{ K})= 1$ ps at all temperatures. In reality, however, the transport lifetime increases strongly with decreasing temperature~\cite{koralek} such that $b_F\,\tau_1\gtrsim 1$ for low temperatures, i.e., the system is outside the D'yakonov-Perel' regime and our theory is no longer applicable. In this low temperature regime other approaches which account for strong spin-orbit interaction could be used.\cite{bernevig3,liu-2011}

\section{Conclusions\label{sec:conclusions}}

Using a spin-coherent Boltzmann-type approach we have derived semiclassical spin-diffusion equations for a two-dimensional electron gas with Rashba and Dresselhaus spin-orbit interactions including the effect of cubic Dresselhaus and extrinsic spin orbit interactions as well as  the influence of electron-electron interactions. Based on this approach we have analyzed the role of electron-electron interaction in generating a finite lifetime of the persistent spin helix state.

Our calculation shows that the Hamiltonian has to contain SU(2)-breaking terms such as cubic Dresselhaus or extrinsic spin orbit interactions in addition to electron-electron interactions. Otherwise the persistent spin helix remains infinitely long-lived. We find that in this respect the effect of extrinsic spin-orbit interaction is negligible in the quantum wells used in the experiment by Koralek \etal\cite{koralek} Instead, the experimentally observed temperature dependence of the lifetime of the persistent spin helix, which displays a maximum at intermediate temperatures close to 100 K, is caused by the interplay of cubic Dresselhaus spin-orbit interaction and electron-electron interactions. The formation of the maximum can be understood as follows: due to electron-electron interactions the scattering rates of the winding number $\pm 3$ components of the spin distribution function grow with increasing temperature. Since the inverse of these rates enters the effective scattering rate in the D'yakonov-Perel' regime, electron-electron interactions increase the PSH lifetime with temperature. On the other hand, Sommerfeld corrections of the cubic Dresselhaus spin orbit interaction enter directly into the expressions for the effective scattering rates and thus decrease the lifetime of the PSH state with increasing temperature. Also temperature-induced deviations from the SU(2) point due to a renormalization of the linear Dresselhaus coupling constant by cubic Dresselhaus spin-orbit interaction increase with temperature and thus effectively reduce the lifetime of the PSH state. Since these corrections due to cubic Dresselhaus spin-orbit interaction dominate for larger temperatures, whereas the effect of electron-electron interaction prevails for lower temperatures, a maximum of the PSH lifetime emerges at intermediate temperatures. 

Our theory reproduces qualitatively  the lifetime of the PSH state in the whole temperature range accessed experimentally by Koralek \etal\cite{koralek}. For intermediate temperatures close to the maximum, i.e., in the regime where our diffusive  theory should be valid, we find also quantitative agreement with the experimental data.

In order to maximize the lifetime, we propose to use a spatially damped sinusoidal spin profile as an initial condition for a transient spin grating spectroscopy experiment. When cubic Dresselhaus spin orbit interaction represents the only SU(2) symmetry breaking element, the proposed spin density profile is infinitely long lived similar to the PSH state in the absence of symmetry breaking terms.

It may be interesting to include also disorder in the local Rashba spin-orbit coupling or spin-dependent electron-electron scattering in order to apply our theory to situations where the cubic Dresselhaus spin-orbit interaction is less dominant. These relaxation mechanisms are currently discussed in the  context of spin relaxation in (110) grown GaAs quantum wells.\cite{sherman, glazov}

\begin{acknowledgments}

We thank F.\ von Oppen for helpful discussions and J.\,D.\ Koralek for providing with experimental data. This work was supported by SPP 1285 of the DFG.
\end{acknowledgments}

\begin{appendix}
\section{Sommerfeld functions}\label{app:sommerfeld}

From the standard Sommerfeld technique in the theory of the Fermi gas it is well known 
that the approximation
\begin{align}
\int_0^\infty d \epsilon\,g(\epsilon)\, f(\epsilon)&~=~\int_0^{E_F} d \epsilon\, g(\epsilon)+\frac{\pi^2}{6} \,(k_B T)^2\,g'(E_F)\nonumber\\
&\qquad+\mathcal{O}(T^4/T_F^4)
\end{align}
holds, where $f(\epsilon)$ is the Fermi distribution and $g(\epsilon)$ is a function of the energy that varies slowly for $\epsilon\approx E_F $. 
 In the derivation of the spin diffusion equations we have to deal with powers of momentum $k^2,k^3,k^4,k^6,k^8$. Since the dispersion is quadratic and the 2d DOS is constant, the problem  reduces to ($n=1,\frac{3}{2},2,3,4$)
\begin{align}
\int_0^\infty d \epsilon\,\epsilon^n\, f'(\epsilon)&= -\int_0^\infty d \epsilon\,n\,\epsilon^{n-1}\, f(\epsilon)\nonumber\\
&= -(E_F)^n\left[1+n\,\left(n-1\right)\frac{\pi^2}{6}\left(\frac{k_B\,T}{E_F}\right)^2\right]\nonumber\\ &\quad +\mathcal{O}(T^4/T_F^4).
\end{align}
Thus, $k^2$ terms do not obtain any $T$-dependent corrections, whereas the higher powers, $k^3,k^4,k^6$ and $k^8$, are not simply replaced by $-k_F^3,\dots-k_F^8$ but acquire corrections in the form of the factors $z_3,\dots z_8$, see Eqs.~\eqref{z3}-\eqref{z8}.

\end{appendix}

\pagebreak


\begin{thebibliography}{xxxxxxx}

\bibitem{book_awschalom_2002} {\em Semiconductor Spintronics and Quantum Computation}, edited by D.\,D.\,Awschalom, D.\,Loss, and N.\,Samarth (Springer, Berlin, 2002).

\bibitem{zutic_2004} I.\,\ifmmode \check{Z}\else \v{Z}\fi{}uti\ifmmode \acute{c}\else \'{c}\fi{}, J.\, Fabian, and S.\, Das Sarma, Rev.\,Mod.\,Phys.\,{\bf 76}, 323 (2004).

\bibitem{awschalom_2007} D.\,D.\,Awschalom and M.\,E.\ Flatt\'e, Nature Phys.\, {\bf 3}, 153 (2007).

\bibitem{awschalom_2009} D.\,D.\,Awschalom and N.\,Samarth, Physics {\bf 2}, 50 (2009). 

\bibitem{schliemann} J.\,Schliemann, J.\,C.\,Egues, and D.\,Loss, Phys.\,Rev.\,Lett.\,{\bf 90}, 146801 (2003).

\bibitem{bernevig1} B.\,A.\,Bernevig, J.\,Orenstein, and S.-C.\,Zhang, Phys.\,Rev.\,Lett.\,{\bf 97}, 236601 (2006).

\bibitem{koralek} J.\,D.\,Koralek, C.\,P.\,Weber, J.\,Orenstein, B.\,A. Bernevig, S.-C.\,Zhang, S.\,Mack, and D.\,D. Awschalom, Nature {\bf 458}, 610 (2009).

\bibitem{cameron} A.\,R.\,Cameron, P.\,Riblet, and A.\,Miller, Phys.\,Rev.\,Lett.\,{\bf 76}, 4793 (1996).

\bibitem{Amico1} I.\ D'Amico and G.\ Vignale, Phys.\ Rev. B  \ {\textbf 62}, 4855 (2000).

\bibitem{flensberg} K.\,Flensberg, T.\,S.\,Jensen, and N.\,A.\,Mortensen, Phys.\,Rev.\,B {\bf 64}, 245308 (2001).

\bibitem{Amico2} I.\,D'Amico and G.\,Vignale, Phys.\,Rev.\,B {\bf 68},  045307 (2003).

\bibitem{weber} C.\,P.\,Weber, N.\,Gedik, J.\,E.\,Moore, J.\,Orenstein, J.\,Stephens, and D.\,D.\,Awschalom, Nature {\bf 437}, 1330 (2005).

\bibitem{raimondi} R.\,Raimondi and P.\,Schwab, Physica E {\bf 42}, 952 (2009).

\bibitem{stanescu} T.\,D.\,Stanescu and V.\,Galitski, Phys.\,Rev.\,B {\bf 75},  125307 (2007).

\bibitem{winkler} R.\,Winkler, {\em Spin-Orbit Coupling Effects in Two-Dimensional Electron and Hole Systems}, Springer Tracts in Modern Physics {\bf 191} (2003).

\bibitem{rashba} Y.\,A.\,Bychkov and E.\,I.\,Rashba, J.\,Phys.\,Chem.\,{\bf 17}, 6039 (1984).

\bibitem{dresselhaus} G.\,Dresselhaus, Phys.\,Rev.\,{\bf 100}, 580 (1955).

\bibitem{weng} M.\,Q.\,Weng, M.\,W.\, Wu, and H.\,L.\,Cui, J.\,Appl.\,Phys.\,{ \bf 103}, 063714 (2008).

\bibitem{AltDerivations} Alternative derivations with the Keldysh formalism or standard density matrix approaches should, to the desired zeroth order in $b_F/E_F$, yield the same equations. Note, however, that to general orders in $b_F/E_F$, important differences between the formalisms may arise, see~Ref.~\onlinecite{kailasvuori}.

\bibitem{kailasvuori} J.\,Kailasvuori and M.\,C.\,L\"uffe, J.\,Stat.\,Mech.\,(2010)\,P06024.


\bibitem{zubarev} D.\,N.\,Zubarev, V.\,Morozov, and G.\,R\"opke, {\it Statistical Mechanics of Nonequilibrium Processes}, Vol. 1 (Akademie Verlag, Berlin, 1996).
                                                                                                                                                                 \bibitem{HartreeFock} In principle, an additional Hartree-Fock precession term (second order in $\vc{s}_\vc{k}$ and first-order in the electron-electron interaction $V$) on the left-hand side of Eq.~\eqref{spin} could become important,\cite{stich} as well as quadratic in $\vc{s}_\vc{k}$ terms in the collision integrals. 
However, for small polarization these effects can be neglected.
The clean double-exponential decay of the transient spin grating as documented in Fig.~1a of Ref.~\onlinecite{koralek} is a hint that in this particular experiment nonlinear effects are irrelevant.

\bibitem{stich} D.\,Stich, J.\,Zhou, T.\,Korn, R.\,Schulz, D.\,Schuh, W.\,Wegschneider, M.\,W.\, Wu, and C.\,Sch{\"u}ller, Phys.\,Rev.\,Lett. {\bf 98}, 176401 (2007).

\bibitem{burkov} A.\,A.\,Burkov, A.S.\,N\'u\~nez, and A.H.\,MacDonald, Phys.\,Rev. B {\bf 70}, 155308 (2004).

\bibitem{mishchenko} E.\,G.\,Mishchenko, A.\,V.\,Shytov, and B.\,I.\, Halperin, Phys.\,Rev.\,Lett.\,{\bf 93}, 226602 (2004).

\bibitem{dyakonov} M.\,I.\,D'yakonov and V.\,I.\ Perel',  Sov.\,Phys.\,Solid State, Vol.\,{\bf 13}, No.\,12 (1972).

\bibitem{yang} L.\,Yang, J.\,Orenstein, and D.-H.\,Lee, Phys.\,Rev.\,B {\bf 82}, 155324 (2010).

\bibitem{lifshits} M.\,B.\,Lifshits and M.\,I.\,D'yakonov', Phys.\,Rev.\,Lett.\,{\bf 103}, 186601 (2009).

\bibitem{bernevig3}  B.\,A.\,Bernevig and J.\,Hu, Phys.\,Rev.\,B {\bf 78}, 245123 (2008).

\bibitem{liu-2011} X.\,Liu, X.-J.\,Liu, and J.\,Sinova, arXiv:1102.3170v1\,(2011).

\bibitem{sherman} E.\,Ya.\,Sherman, Appl.\,Phys.\,Lett. {\bf 82}, 2, 209 (2003).

\bibitem{glazov} M.\,M.\,Glazov, M.\,A.\,Semina, and E.\,Ya.\,Sherman, Phys.\,Rev.\,B {\bf 81},  115332 (2010). 

\end{thebibliography}
\end{document}